\documentclass{ws-procs9x6}

\begin{document}

\title{Remarks on the reweighting method in the chemical potential direction
\footnote{\uppercase{T}his work is supported by \uppercase{PPARC} grant 
\uppercase{PPA/G/S/}1999/00026.
}}

\author{Shinji Ejiri 
\footnote{\uppercase{T}his report is based on work for the 
\uppercase{B}ielefeld-\uppercase{S}wansea \uppercase{C}ollaboration.}}

\address{Department of Physics, University of Wales Swansea, \\
Singleton Park, Swansea, SA2 8PP, U.K.}


\maketitle

\abstracts{
We comment on the reweighting method in the chemical potential 
$(\mu_{\rm q})$ direction. We study the fluctuation of the 
reweighting factor during Monte-Carlo steps. 
We find that it is the absolute value of the reweighting factor that 
mainly contributes to the shift of the phase transition line $(\beta_c)$ 
by the presence of $\mu_{\rm q}$. 
The phase fluctuation is a cause of the sign problem, 
but the effect on $\beta_c$ seems to be small. We also discuss $\beta_c$ 
for Iso-vector chemical potential and $\beta_c$ determined from 
simulations with imaginary chemical potential.}

\section{Introduction}
\label{sec:intro}

The study of QCD at finite temperature and finite density is currently one 
of the most attractive topics in particle physics. 
The heavy-ion collision experiments aiming to produce the 
quark-gluon plasma are running at BNL and CERN, 
for which the interesting regime is rather low density.  
Moreover a new color superconductor phase is expected 
in the region of low temperature and high density. 
Numerical study by Monte-Carlo simulations of Lattice QCD is a 
powerful means of investigating aspects of the phase transition 
but the Monte-Carlo method is not applicable directly at finite density 
because the fermion determinant is complex for non-zero quark chemical 
potential $\mu_q$. Most of the studies at $\mu_{\rm q} \neq 0$ are done 
by the reweighting method performing simulations at 
${\rm Re}(\mu_{\rm q})=0$\cite{Bar97,Fod01,swan02}. 

In this paper, we make a comment on the reweighting method for 
the chemical potential direction. The interesting point is that 
there are two simulation parameters, $\beta=6/g^2$ and $\mu=\mu_q a$. 
We perform a simulation at a suitable point $(\beta_0, \mu_0)$ and, 
in order to calculate an observable at another point $(\beta, \mu)$, 
we modify the Boltzmann weight.
Then, if the modification factors (reweighting factors) for $\beta$ and 
$\mu$ are correlated during Monte-Carlo steps, there could exist 
a parameter subspace in which the effect of the modification is 
cancelled and changes of all physical quantities are small. 
Therefore it is interesting to investigate the correlation between 
the reweighting factors and to confirm if such a direction exists. 
It is also interesting to discuss the relation between that direction 
and the phase transition line, since we naively expect the physics 
is similar along the phase transition line. Moreover increase of the error 
due to the reweighting could be reduced by the cancellation, 
which may explain why Fodor and Katz\cite{Fod01} could calculate 
$\beta_c$ for rather large $\mu$.

\section{Fluctuation of the reweighting factor}
\label{sec:method}

The reweighting method is based on the following identity:
\begin{eqnarray}
\langle O \rangle_{(\beta, \mu)}
&=& \frac{1}{Z_{(\beta, \mu)}} \int 
D U O ( \det M_{(\mu)})^{N_{\rm f}} e^{-S_g(\beta)} 
\label{eq:rew} 
= \frac{\left\langle O e^{\Delta F} e^{\Delta G}
 \right\rangle_{(\beta_0, \mu_0)} }{\left\langle e^{\Delta F} 
e^{\Delta G}  \right\rangle_{(\beta_0, \mu_0)}}. 
\end{eqnarray}
Here $M$ is the quark matrix, $S_g$ is the gauge action, 
$N_{\rm f}$ is the number of flavors, 
$F=N_{\rm f}(\ln \det M_{(\mu)}-\ln \det M_{(\mu_0)}),$
$G=(\beta-\beta_0) P, P=-\partial S_g/\partial \beta,$ and 
$\Delta f = f - \langle f \rangle$.
The expectation value $\langle O \rangle_{(\beta, \mu)}$ can in principle 
be computed by simulation at $(\beta_0, \mu_0)$ by this identity.
In this study, we put $\mu_0=0$.
If $O$ and 
$e^{\Delta F} e^{\Delta G} =e^F e^G /(e^{\langle F \rangle} 
e^{\langle G \rangle})$ fluctuate with a correlation 
during Monte-Carlo steps, $\langle O \rangle_{(\beta, \mu)}$ has 
$\mu$ and $\beta$ dependence. 
Otherwise, the $\mu$ and $\beta$ dependence cannot be obtained, 
e.g. if $ e^F e^G$ does not fluctuate, 
$ e^{\Delta F} e^{\Delta G}=1$ and $\langle O \rangle_{(\beta, \mu)}$ 
does not change.
Roughly speaking, the difference of $\langle O \rangle_{(\beta, \mu)}$ 
from $\langle O \rangle_{(\beta_0, 0)}$ increases as the magnitude of 
fluctuations of $F$ and $G$ increases.

We discuss the correlation between $e^{F}$ and $ e^{G}$. 
Since $e^{F}$ is complex, we separate it into a phase 
factor and an amplitude. 
As is shown in Ref.\cite{swan02}, the phase factor and the amplitude 
can be written by the odd terms and the even terms of the Taylor 
expansion of $\ln \det M$, respectively, since the odd terms are 
purely imaginary and the even terms are real at $\mu=0$. 
Denoting $e^{F} = e^{i \theta} |e^{F}|$ and 
$F=\sum_{n=1}^{\infty} R_n \mu^n,$ 
$|e^{F}|=\exp\{\sum_{n=1}^{\infty} {\rm Re} 
R_{2n} \mu^{2n}\},$ and 
$e^{i \theta}=\exp\{i \sum_{n=1}^{\infty} {\rm Im} 
R_{2n-1} \mu^{2n-1}\}.$
We study these correlations in the vicinity of the simulation point 
$(\beta_0, 0)$. Up to $O[\beta-\beta_0, \mu^2]$, the reweighting factor is 
$ e^{i \theta} |e^{F}| e^{G} \approx 1 + R_1 \mu 
+ R_1^2 \mu^2 /2 + R_2 \mu^2 + P (\beta-\beta_0)$. 
We compute the correlations, 
$\langle \Delta (R_1^2/2) \Delta P \rangle,$
$\langle \Delta R_2 \Delta P \rangle,$ and 
$\langle \Delta (R_1^2/2) \Delta R_2 \rangle,$ 
which correspond to the correlations of $(e^{i \theta}, e^{G})$, 
$(|e^{F}|, e^{G})$ and $(e^{i \theta}, |e^{F}|)$, 
respectively. Here, $\langle \Delta R_1 \Delta P \rangle$ is zero 
at $\mu=0$ because $R_1$ is purely imaginary.

We use the configurations in Ref.\cite{swan02}, which are generated 
by the $N_{\rm f}=2$ $p4$-improved staggered action on a 
$16^3 \times 4$ lattice. 
The results are summarized in Table~\ref{tab1}. 
We find that the correlation between $|e^{F}|$ and 
$e^{G}$ is very strong in comparison with the other 
correlations, which means that the contribution to an observable can be 
separated into two independent parts: from $e^{i \theta}$, and 
from a combination of $|e^{F}| \times e^{G}.$

\begin{table}[ht]
\tbl{Correlations among $R_1^2, R_2,$ and $P$. 
$N_{\rm site} =16^3 \times 4.$
\vspace*{1pt}}
{\footnotesize
\begin{tabular}{|ccccc|}
\hline
{} &{} &{} &{} &{}\\[-1.5ex]
$m$ & $\beta$ & 
$\langle \Delta (R_1^2/2) \Delta P \rangle N_{\rm site}^{-1}$ & 
$\langle \Delta R_2 \Delta P \rangle N_{\rm site}^{-1}$ & 
$\langle \Delta (R_1^2/2) \Delta R_2 \rangle N_{\rm site}^{-1}$\\[1ex]
0.1 &3.64 &0.006(29) &0.312(33) &0.034(10)\\[1ex]
{}  &3.65 &0.059(21) &0.434(29) &0.056(10)\\[1ex]
{}  &3.66 &0.055(15) &0.410(26) &0.022(5)\\[1ex]
{}  &3.67 &0.032(15) &0.397(28) &0.031(5)\\[1ex]
0.2 &3.75 &0.037(18) &0.287(26) &0.029(7)\\[1ex]
{}  &3.76 &0.019(10) &0.353(23) &0.018(3)\\[1ex]
{}  &3.77 &0.037(10) &0.359(24) &0.017(3)\\[1ex]
\hline
\end{tabular}\label{tab1} }
\vspace*{-10pt}
\end{table}

To make the meaning of this result clearer, we consider the following 
partition function, introducing two different $\mu$, $\mu_o$ and $\mu_e$, 
\begin{eqnarray}
\label{eq:par}
Z= \int D U e^{R_1 \mu_o +R_3 \mu_o^3 +\cdots} 
e^{R_2 \mu_e^2 +R_4 \mu_e^4 +\cdots} 
(\det M|_{\mu=0})^{N_{\rm f}} e^{-S_g} 
\label{eq:modact} 
\end{eqnarray}
Then, 
$\langle \Delta R_1^2 \Delta P \rangle 
= \langle (\Delta R_1)^2 \Delta P \rangle 
= \frac{\partial^3 \ln Z}{\partial \mu_o^2 \partial \beta} 
= \frac{\partial (\chi_{\rm S} - \chi_{\rm NS})}{\partial \beta}
 N_{\rm site},$ 
$2 \langle \Delta R_2 \Delta P \rangle 
= \frac{\partial^3 \ln Z}{\partial \mu_e^2 \partial \beta} 
= \frac{\partial \chi_{\rm NS}}{\partial \beta} N_{\rm site}, $
at $\mu=0$, 
where $\chi_{\rm S}$ and $\chi_{\rm NS}$ are the singlet and non-singlet 
quark number susceptibilities\cite{Got88}.
The result in Table~\ref{tab1} means that 
$\frac{\partial^3 \ln Z}{\partial \mu_o^2 \partial \beta} 
 << \frac{\partial^3 \ln Z}{\partial \mu_e^2 
\partial \beta},$
i.e. $\mu$ in the phase factor $(\mu_o)$ does not contribute to the 
$\beta$-dependence of $Z$ and $\mu$ in the amplitude $(\mu_e)$ 
is more important for the determination of $\beta_c$.
Moreover, these correlations have a relation with the slope of 
$\chi_{\rm S}$ and $\chi_{\rm NS}$ in terms of $\beta$. 
Since $\chi_{\rm S} - \chi_{\rm NS}$ is known to be small\cite{Got88}, 
this result may not change even for small quark mass.

\paragraph{Iso-vector chemical potential} 
Next, we discuss the model with Iso-vector chemical potential\cite{Son01}. 
If we impose a chemical potential with opposite sign for $u$ and 
$d$ quarks: $\mu_u=-\mu_d,$ the Monte-Carlo method is applicable, 
since the measure is not complex. 
In Ref.\cite{swan02}, we discussed the difference of the curvature 
of the phase transition line in this case.
Because we expect at $T=0$ that pion condensation happens around 
$\mu_{\rm q} \approx m_{\pi}/2$, if we consider that the phase transition line 
runs to that point directly, the curvature of the transition line for 
Iso-vector $\mu$ should be much larger than that for usual $\mu$, 
since $m_{\pi}/2 << m_{\rm N}/3$. 
However, as we discussed above, $\mu_o$ in Eq.~\ref{eq:par} does not 
contribute the shift of $\beta_c$ and the difference from the usual 
$\mu$ is only in $\mu_o$, i.e. $\mu_o=0$ for the Iso-vector case. 
Therefore the difference of the curvature might be small and 
the naive picture seems to be wrong.
In practice, our result at small $\mu$ using the method in 
Ref.\cite{swan02} supports that. 
Moreover Kogut and Sinclair\cite{Kog02} showed that $\beta_c$ from 
chiral condensate measurements is fairly insensitive to $\mu$ for small 
$\mu$ by direct simulations with Iso-vector $\mu$. 

\begin{figure}[ht]
\centerline{
\epsfxsize=2.0in\epsfbox{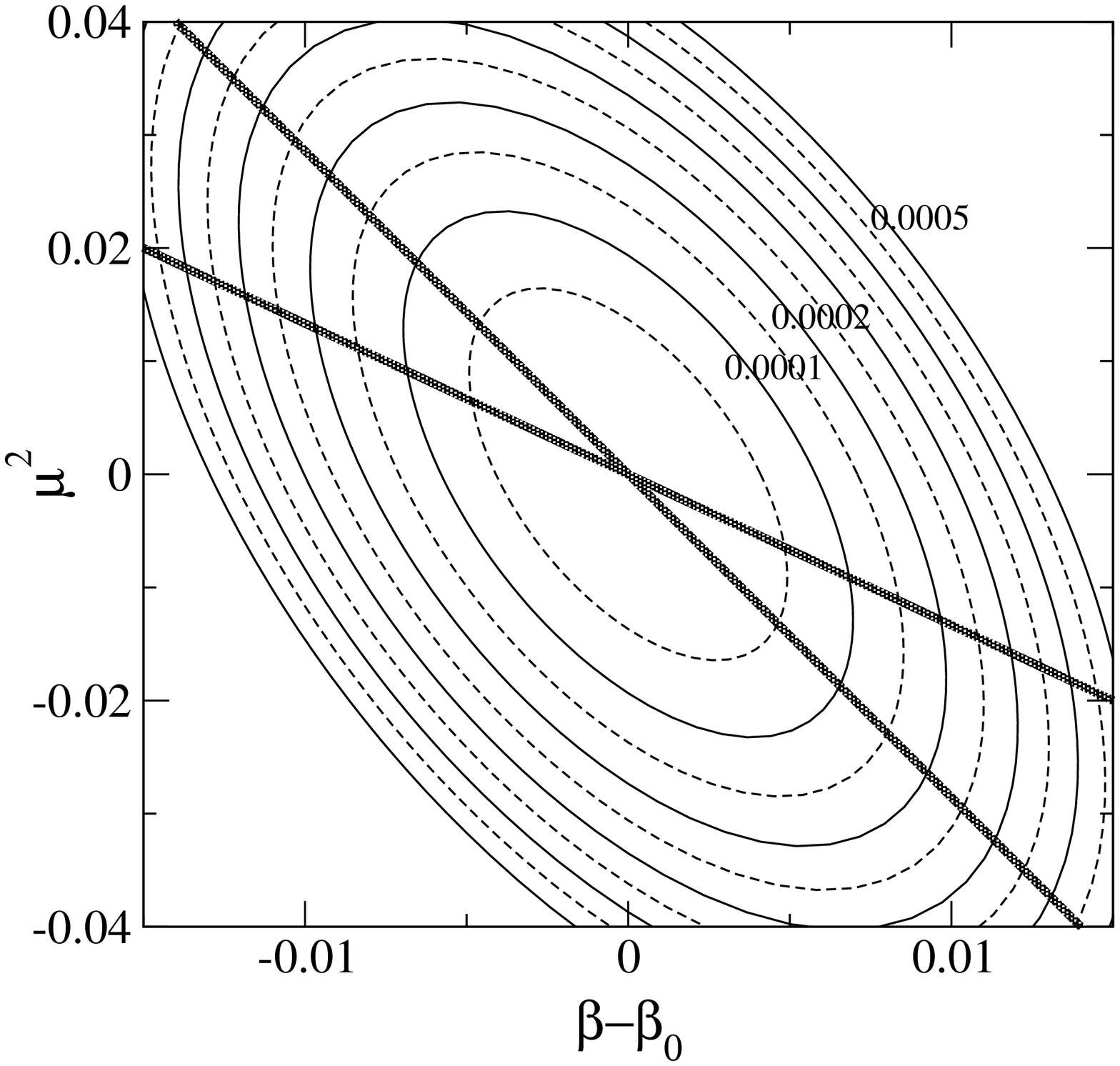}   
\epsfxsize=2.15in\epsfbox{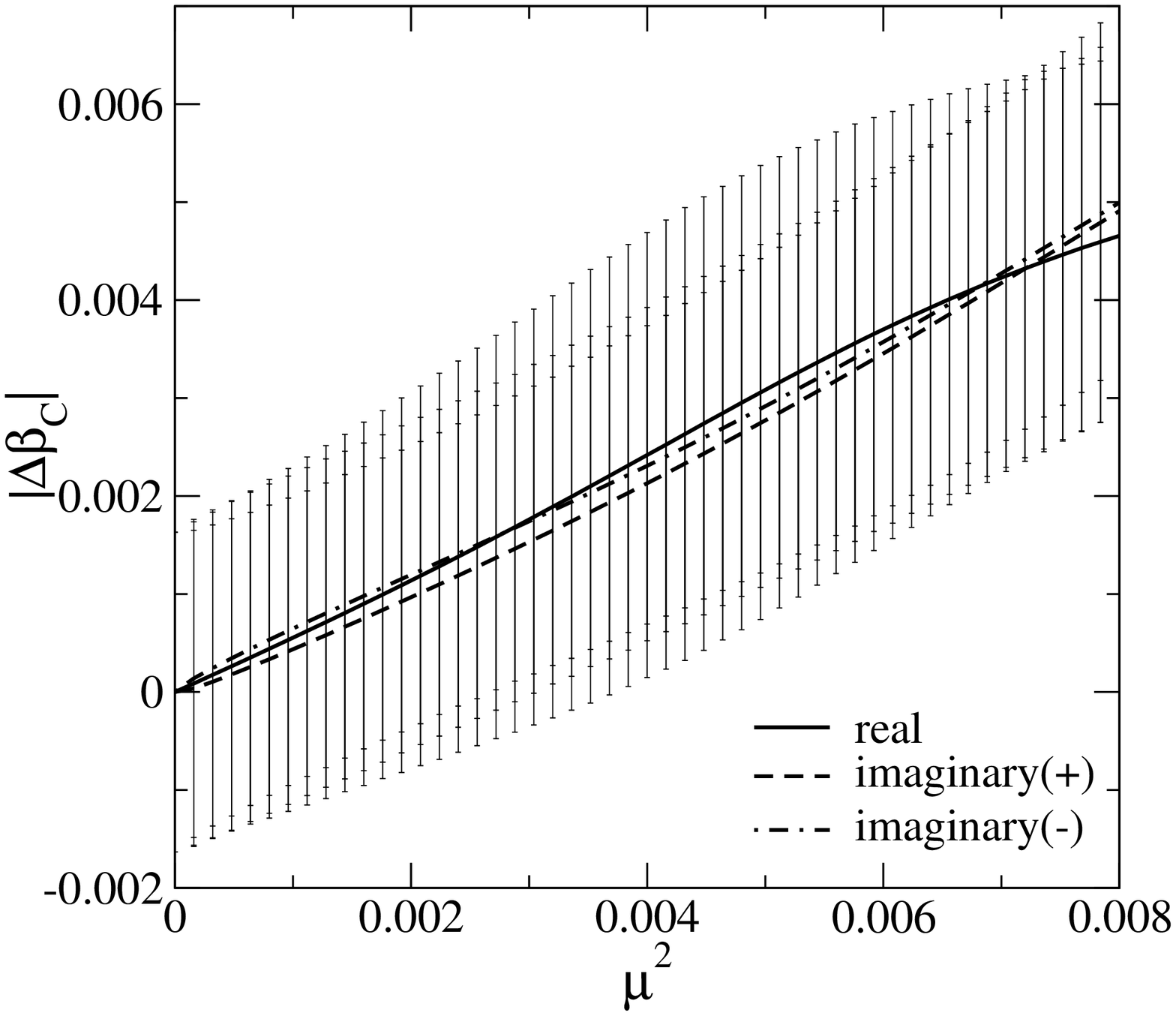}}
\caption{Left figure is the contour plot of the dispersion around 
$\beta_c(0)$. Bold lines show $\partial^2 \beta_c / \partial \mu^2$. 
$\beta_0=3.6497$, $m=0.1$. 
Right figure shows $|\beta_c(\mu)-\beta_c(0)|$ for 
real $\mu$ and imaginary $\mu$ at $m=0.1$. \label{fig1}}   
\end{figure}
\vspace{-2mm}

We estimate the fluctuation of the reweighting factor. 
The phase factor $e^{i \theta}$ gives a contribution independent 
from the other parts. 
As we discussed in Ref.\cite{swan02}, these phase fluctuations cause 
the sign problem to become more severe as $\mu$ 
(or $\mu_o$) increases. The amplitude of the fermionic part and 
the gauge part are strongly correlated. 
Therefore the magnitude of the contribution to a observable is not 
simple. We compute the dispersion of $|e^{F}| e^{G}$ 
to estimate the fluctuation. Up to $O[\beta-\beta_0, \mu^2]$, 
$\langle \{\Delta (|e^{F}| e^{G}) \}^2 \rangle \approx 
\mu^4 \langle (\Delta R_2)^2 \rangle 
+2\mu^2 (\beta-\beta_0) \langle \Delta R_2 \Delta P \rangle
+(\beta-\beta_0)^2 \langle (\Delta P)^2 \rangle$.
Then, the line of constant dispersion is an ellipse.
We write contour line in Fig.~\ref{fig1}(left).
Here the susceptibilities of $R_2$ and $P$ and the correlation of 
$R_2$ and $P$ are computed at the phase transition point, 
$\beta_c=3.6497(16)$ for $m=0.1$.   
We denote the lower and upper bounds of 
$\partial^2 \beta_c / \partial \mu^2=-1.1(4)$ by bold lines\cite{swan02}.
We find that there exists a direction along which the increase of 
the fluctuation is relatively small. 
That direction is roughly parallel to the phase transition line.
Because we expect that physics is similar along the transition line, 
if we consider that $|e^{F}| e^{G}$ is the important part 
for the calculation of $\beta_c$, this result suggests that 
the phase transition line is determined by the quite simple mechanism 
that the fluctuation of the reweighting factor itself is small 
along the transition line and physics is similar on that line.

\paragraph{Imaginary chemical potential}
In Fig.~\ref{fig1}(left), we write also the region for $\mu^2<0$, 
i.e. imaginary $\mu$. 
de Forcrand and Philipsen\cite{For02} computed 
$\partial^2 \beta_c / \partial \mu^2$ performing simulations with 
imaginary $\mu$ , assuming that $\beta_c$ is an even function in 
$\mu$ and analyticity in that region. 
Here, we confirm whether the results obtained by real and imaginary 
$\mu$ are consistent or not by the method in Ref.\cite{swan02}.  
We replace $\mu$ by $i\mu$ or $-i\mu$ and reanalyze for imaginary $\mu$. 
In Ref.\cite{swan02}, the reweighting factor has been obtained 
in the form of the Taylor expansion in $\mu$ up to $O(\mu^2)$, 
and the replacement is easy. 
We determined $\beta_c$ by the chiral susceptibility. 
The results of $|\beta_c(\mu) -\beta_c(0)|$ are written 
in Fig.~\ref{fig1}(right). Errors are $O(\mu^4)$.
The solid line is the result for real $\mu$. 
The results of $\mu \to i\mu$ and $\mu \to -i\mu$ are dashed and 
dot-dashed lines respectively. 
The slope at $\mu=0$ is $(\partial^2 \beta_c / \partial \mu^2)/2$. 
We find that these results of the slope for real and imaginary $\mu$ 
are consistent.

\section{Conclusions}

We investigated the fluctuation of the reweighting factor. 
The contribution to the $(\beta, \mu)$ dependence of a physical 
quantity can be separated into the phase factor and the amplitude. 
Mainly the amplitude of the reweighting factor contributes to 
the determination of $\beta_c$. 
The fluctuation of the amplitude is small along the phase 
transition line, which is consistent with physics being similar 
along the transition line,
while the fluctuation of the phase increases in proportion to $\mu$ and 
causes the sign problem.
We also confirmed that the second derivative of the phase transition 
lines at $\mu=0$ determined from imaginary $\mu$ is consistent with 
that from real $\mu$.

%

\end{document}